\documentclass[conference]{IEEEtran}

\IEEEoverridecommandlockouts                              
\usepackage{cite}
\usepackage{graphicx}
\usepackage{amsmath}
\usepackage{multirow}
\usepackage{flushend}
\usepackage{xcolor}
\newcommand\T{\rule{0pt}{2.6ex}}

\def\BibTeX{{\rm B\kern-.05em{\sc i\kern-.025em b}\kern-.08em
    T\kern-.1667em\lower.7ex\hbox{E}\kern-.125emX}}

\title{A Spanning Tree-based Genetic Algorithm for Distribution Network Reconfiguration}

\author{Mukesh Gautam*, \emph{Student Member, IEEE}, Narayan Bhusal*, \emph{Student Member, IEEE},\\ Mohammed Benidris*, \emph{Member, IEEE}, and Sushil J. Louis**, \emph{Member, IEEE}, \\
*Department of Electrical and Biomedical Engineering, 
\\ **Department of Computer Science and Engineering,\\  
University of Nevada, Reno, NV 89557, USA\\
Emails: mukesh.gautam@nevada.unr.edu, bhusalnarayan62@nevada.unr.edu, \\  mbenidris@unr.edu, and sushil@cse.unr.edu}

\begin{document}
\maketitle
\thispagestyle{empty}
\pagestyle{empty}

\begin{abstract}
This paper presents a spanning tree-based genetic algorithm (GA) for the reconfiguration of electrical distribution systems with the objective of minimizing active power losses. Due to low voltage levels at distribution systems, power losses are high and sensitive to system configuration. Therefore, optimal reconfiguration is an important factor in the operation of distribution systems to minimize active power losses. Smart and automated electric distribution systems are able to reconfigure as a response to changes in load levels to minimize active power losses. The proposed method searches spanning trees of potential configurations and finds the optimal spanning tree using genetic algorithm in two steps.  In the first step, all invalid combinations of branches and tie-lines (i.e., switching combinations that do not provide power to some of loads or violate the radiality and connectivity conditions) generated by initial population of GA are filtered out with the help of spanning tree search algorithm. In the second step, power flow analyses are performed only for combinations that form spanning trees. The optimal configuration is then determined based on the amount of active power losses (optimal configuration is the one that results in minimum power losses). The proposed method is implemented on several systems including the well-known 33-node and 69-node systems. The results show that the proposed method is accurate and efficient in comparison with existing methods.   
\end{abstract}
\section{Introduction}
Power losses minimization is an important factor in the operation of electrical distribution systems. Electrical distribution systems are characterized by having high resistance-to-reactance ratio, low voltage levels, and radial or weakly-meshed structures.  Active power losses ($I^2R$ loss) is high due to high branch resistances and low voltage level (i.e., requires high current flows). Therefore, it is critical to develop methods for optimal reconfiguration to minimize power losses. Although this is a well-known and amply studied problem, many technical and mathematical problems still exist. Spanning tree based genetic algorithm methods have the potential to find optimal or near-optimal solutions for this type of problems. 

Distribution network reconfiguration (DNR) is an important real-time operation task which is performed to achieve various objectives (e.g., power loss minimization, reliability improvement, maximum load restoration, voltage deviation minimization, load balancing, peak shaving, and operational cost minimization). Distribution systems are generally  equipped  with  two  types of  switches:  sectionalizing  switches  (normally  closed)  and tie-switches  (normally  open)  to  serve  the  maximum  loads during  the  normal  and  the  contingency  conditions.  The DNR is the process of modifying the configuration of distribution networks through changing statuses of sectionalizing and tie switches to achieve targeted objective(s) while satisfying system constraints (operational and technical). Therefore, DNR can be viewed as an optimization problem. The simplest method of solving the DNR problem is to exhaustively search for all possible combinations. However, exhaustive search techniques are not computationally attractive in the case of electrical distribution systems because of the large search space and the dynamic nature of electrical loads. Numerous analytical and population-based intelligent search techniques (e.g., branch and bound methods, expert systems, and evolutionary algorithms) have been developed and used in literature to solve DNR problems.  

The early work on DNR dates back to 1975 when Merlin and Back \cite{MerlinandBack} first proposed branch and bound method for reducing active power losses in distribution networks. An algorithm to determine switching decisions for DNR has been proposed in \cite{4113836}. A reconfiguration algorithm based on the interaction between service restoration and load balance has been proposed in \cite{CASTRO1985155}. A heuristic algorithm based on branch-and-bound strategy has been employed in \cite{6331584} to solve the network reconfiguration problem for minimum power loss and maximum network current. In \cite{4302553}, the authors have proposed a network reconfiguration algorithm based on a fuzzy multi-objective approach. A two-stage sequential Monte Carlo simulation (MCS)-based stochastic strategy has been proposed in \cite{PES2020NB} to determine the minimum size of movable energy resources (MERs) for service restoration and reliability enhancement. In the approach proposed in \cite{PES2020NB}, the authors have incorporated a spanning tree search algorithm for optimal network configuration, Dijkstra's shortest path algorithm for optimal routes to deploy MERs, and the traveling time of MERs for distribution systems optimal operation. A meta-heuristic Harmony Search Algorithm (HSA) has been used in \cite{6205640} for optimal reconfiguration and Distributed Generation (DG) placement in distribution systems. A new rule-based codification for various meta-heuristic techniques has been proposed in \cite{SWARNKAR20111619} to solve the reconfiguration problems of distribution network. A modified form of particle swarm optimization (PSO) has been proposed in \cite{8316173} for effective identification of the optimal configuration of distribution network. The online reconfiguration of active distribution networks for maximum integration of DG has been proposed in \cite{7776824}. 

Although numerous analytical and population-based methods have been proposed in literature for solving the DNR problem, spanning tree based methods have been given little attention. The authors in \cite{6344718} have proposed minimum spanning tree using Kruskal's algorithm for distribution network reconfiguration. In our paper, both spanning tree search algorithm and genetic algorithm are combined for solving DNR problem.

In this paper, a spanning tree based genetic algorithm (GA) is proposed for the reconfiguration of electrical distribution systems for minimum active power losses. 
The proposed method starts by initializing a population of randomly generated configurations. After performing crossover and mutation operations, the population is passed through selection process where each of the individuals is evaluated in two steps.
In first step, all invalid combinations of branches and tie-lines in each chromosome of GA are filtered out with the help of spanning tree search algorithm. In the second step, power flow analyses are performed for only those combinations that form spanning trees and the optimal configuration is determined based on the amount of active power losses (optimal configuration is one with minimum power losses).
The proposed method is implemented on several test systems including the $33$-node and $69$-node distribution systems. 

The remainder of the paper is organized as follows. Section \ref{prob} describes the problem of distribution network reconfiguration. Section \ref{methods} presents the proposed methodology for the implementation of spanning tree based genetic algorithm. Section \ref{results} validates the proposed approach with case studies and discussions. Section \ref{conclusion} provides concluding remarks.

\section{Problem Formulation}\label{prob}
Power loss is an important operational measure which has significant impact on both technical and economic aspects of distribution system operation. Therefore, proper consideration  should be  given  for  minimizing  power  losses in distribution systems.

The total active power loss can be calculated as follows.
\begin{equation} \label{loss}
    P_{loss}=\sum_{k=1}^{E_s} I_k^2 R_k
\end{equation}
In \eqref{loss}, $I_k$ and $R_k$ are, respectively, current and resistance of branch $k$ and $E_s$ is the total number of branches (edges in spanning tree). The power loss minimization function, $P_{loss}$, of \eqref{loss} can be converted into maximization fitness function, $F$, as follows.
\begin{equation} \label{fitness}
    F = \frac{1}{1 + P_{loss}}
\end{equation}
Subject to: 
\begin{gather*}
\sum S_{G,i}-\sum S_{L, i}-P_{loss}=0\mbox{,} \tag{3} \label{equ:gen_load}  \\ S_G^{min} \leq S_G \leq S_G^{max}\mbox{,} \tag{4} \label{equ:gen_limit} \\ V_k^{min}\leq V_{k}\leq V_k^{max}\mbox{,} \tag{5} \label{equ:voltage_limit} \\
\text{Radial topology constraints, } \tag{6}  \label{equ:topology}\\
\text{Node traversing,} \tag{7} \label{equ: load}
\end{gather*}
where \eqref{equ:gen_load} denotes the power balance equation ($S_{G, i}$, $S_{L, i}$, and $P_{loss}$ represents the generation, load, and line loss, respectively); equation \eqref{equ:gen_limit} refers to the generation limits constraint; equation \eqref{equ:voltage_limit} represents voltage limits constraint; equation \eqref{equ:topology} represents topology constraints to maintain the radial topology of the distribution system; and equation \eqref{equ: load} denotes node traversing constraints to supply all loads.

As several combinations of reconfigurations are possible for large networks, it is very challenging to solve this problem. The complexity of the reconfiguration can be presented as follows. 
\begin{equation*} \label{comb}
    C(E,E_s)=\frac{E!}{(E-E_s)! \times E_s!} \tag{8}
\end{equation*}
where $E$ is the total number of edges and $E_s$ is the number of edges in a spanning tree.

If this problem is solved by exhaustively searching all combinations of candidate solutions, the search space will be very large. From \eqref{comb}, it is obvious that the search space increases as the size of network increases. For example, the search space would be $435,897$ and $15,020,334$ respectively for 33-node and 69-node distribution test systems.

As DNR is a large-scale, non-convex, complex, and non-linear problem with a large number of local optima \cite{7867177}, this problem has been solved using various heuristic, meta-heuristic, and artificial intelligence techniques to determine the global optima. These methods not only simplify the problem but also reduce the computation time. Therefore, GA is adopted with spanning tree search algorithm in this paper to solve the DNR problem to minimize active power losses.

\section{Methodology}\label{methods}
This paper proposes a spanning tree based genetic algorithm for solving the DNR problem. The proposed method starts by searching all possible configurations using the spanning tree search algorithm and the optimal reconfiguration that minimizes the power losses which is determined using genetic algorithm. This section describes the spanning tree search algorithm, genetic algorithm, and the proposed solution approach for optimal feeder reconfiguration to reduce active power losses.

\subsection{Spanning Tree Search Algorithm}
For a graph $G$ with $V$ vertices and $E$ edges, a spanning tree can be defined as a subset of the graph, which has a minimum number of edges (say, $E_s$) connecting all vertices (or nodes). The number of edges of a spanning tree $(E_s)$ is less than the number of vertices by $1$. A spanning tree does not have loops and it is not disconnected. A connected graph can have several spanning trees and all possible spanning trees will have same number of edges and nodes. Each of the edges in the graph $G$ has certain values (or weights). The edge weights are problem specific. While determining the minimum spanning tree, the total sum of all edge weights of a spanning tree is minimized.

Electrical distribution networks are composed of nodes and lines similar to vertices and edges of a graph in graph theory. Therefore, an electric distribution network can be viewed as a graph $G$ with $V$ vertices (or nodes) and $E$ edges (or lines). If all nodes are radially connected, they will obviously represent a spanning tree. In case of electric distribution network, edge weights can represent the active power loss of the line. Since the active power loss is proportional to the square of line current, the edge weight in this case changes with the change in network configuration. This is different from the standard graph in which edge weight is assumed to remain constant while determining the minimum spanning tree.
\begin{figure}[b!]
\vspace{-2ex}
    \hspace{-3.2ex}
    \includegraphics[scale=1]{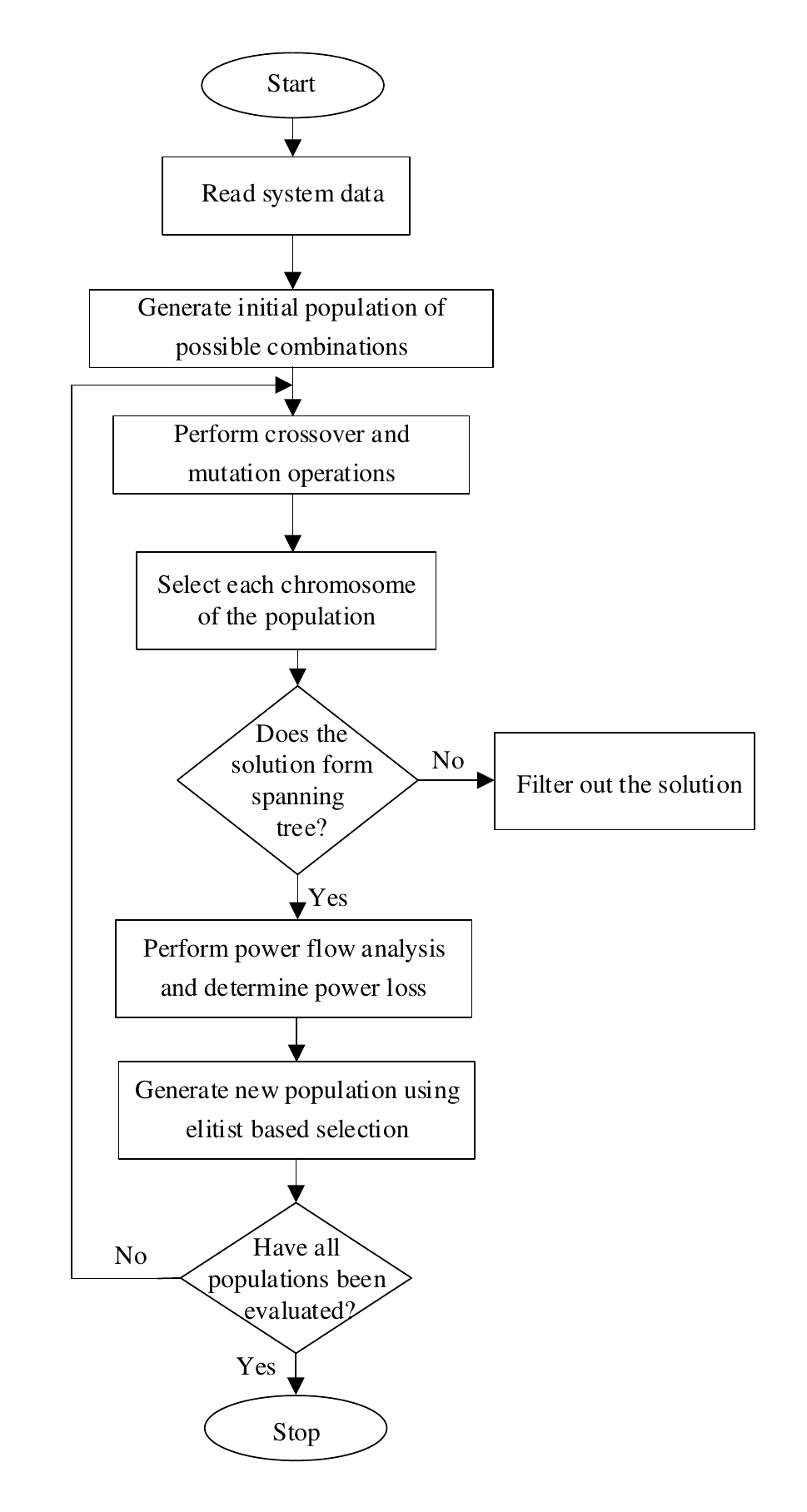}
    \vspace{-5ex}
    \caption{The flowchart of the proposed methodology.}
    \label{flowchart}
    \vspace{-1ex}
\end{figure}

\subsection{Genetic Algorithm}
Genetic algorithm is one of the evolutionary techniques that attempts to mimic some of the processes taking place in natural evolutions. The GA, which is based on Darwin's theory of natural evolution, was first proposed by Holland in 1975 \cite{holland1975}. A GA allows a population composed of many individuals to evolve in such a way that the fitness function is maximized (i.e., the loss is minimized). The fittest individuals are selected for reproduction in GA. The advantages of GA are as follows.
\begin{itemize}
    \item Can optimize both continuous and discrete variables;
    \item Does not require derivative information;
    \item Searches for global optima rather than local optima of even non-convex and non-linear problems; and
    \item Can deal with large number of variables.
\end{itemize}

\subsection{Solution Representation}
Several methods have been used to represent or encode possible combinations of lines and switches in a particular distribution network. Most of previous work in the literature have used binary numbers to represent the status (open/close) of lines and switches. However, in this work, all edges (lines and switches) are numbered from $1$ to total number of edges. Edge numbers are used to generate population of chromosomes because integer number representation reduces the dimensionality of GA search space. For example, in case of the 33-node system, all $37$ edges are numbered from $1$ to $37$ where these numbers are used to generate a string of numbers having string length of $32$.

\subsection{GA Operations}
The GA operations such as cross-over, mutation, and selection used in this work are explained as follows.

\subsubsection{Cross-over}
The partially matched crossover (PMX) is used as a crossover operator in this work. Under PMX, two chromosome strings are aligned, and two cross-over sites are picked randomly along each chromosome string. These two points define a matching section that is used for performing crossover through position-by-position exchange operations \cite{Goldberg1989}.

\subsubsection{Mutation}
In this work, each individual chromosome is composed of $E_s$ number of edges out of total $E$ number of edges. For performing the mutation operation, a random edge from $E_s$ edges is replaced by another random edge from the remaining $(E-E_s)$ edges.

\subsubsection{Elitist Selection}
Elites are the individuals with best fitness in current generation. In the selection methods other than elitist selection, there is a chance of elites being eliminated through cross-over and mutation operations. Therefore, in this work, elitist selection is applied for preserving the best fit individuals, which are automatically passed to the next generation. 

The flowchart of the proposed methodology is shown in Fig \ref{flowchart}.

\begin{figure}
    \centering
    \includegraphics{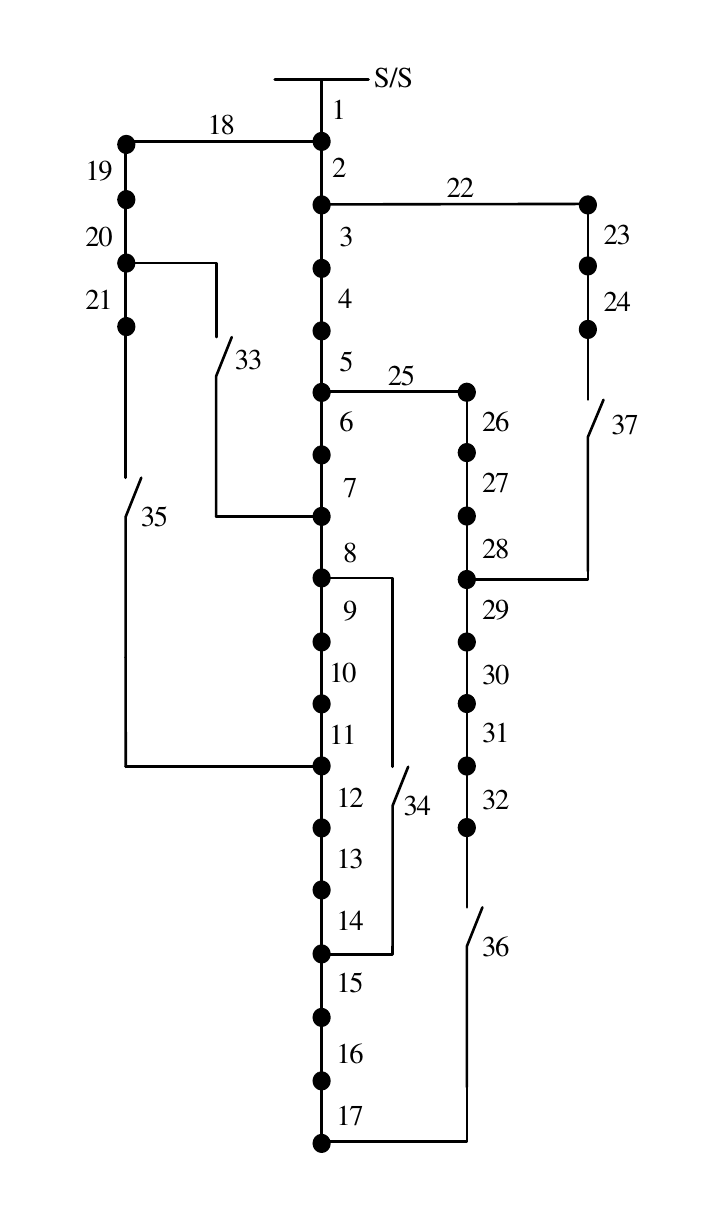}
    \caption{33-node distribution system with five tie-lines before reconfiguration.}
    \label{33bus}
\end{figure}

\section{Case Studies and Discussion}\label{results}
The proposed methodology is implemented on the $33$-node and $69$-node distribution systems. The $33$-node distribution test system is $100$ kVA, $12.66$ kV radial distribution system with $33$ nodes, $32$ branches and $5$ tie-lines \cite{6826880}. Therefore, the total number of branches in this system is $37$. The $33$-node system is shown in Fig. \ref{33bus}, where all the branches (including tie-lines) are numbered from $1$ to $37$.  
\begin{figure}
    \centering
    \includegraphics[scale=1]{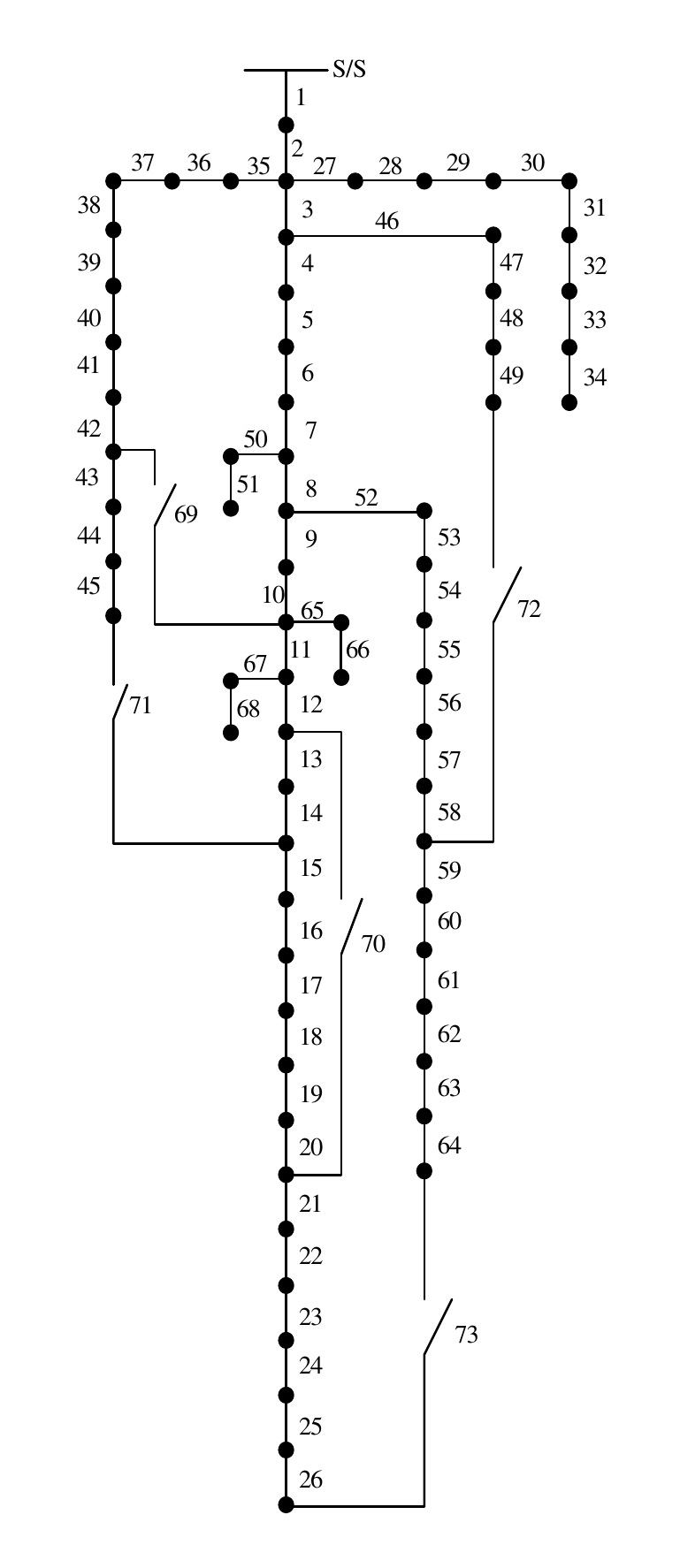}
    \vspace{-2ex}
    \caption{69-node distribution system with five tie-lines before reconfiguration.}
    \vspace{-2ex}
    \label{69bus}
\end{figure}

\begin{figure*}
\hspace{-6ex}
    \includegraphics[scale=1.3]{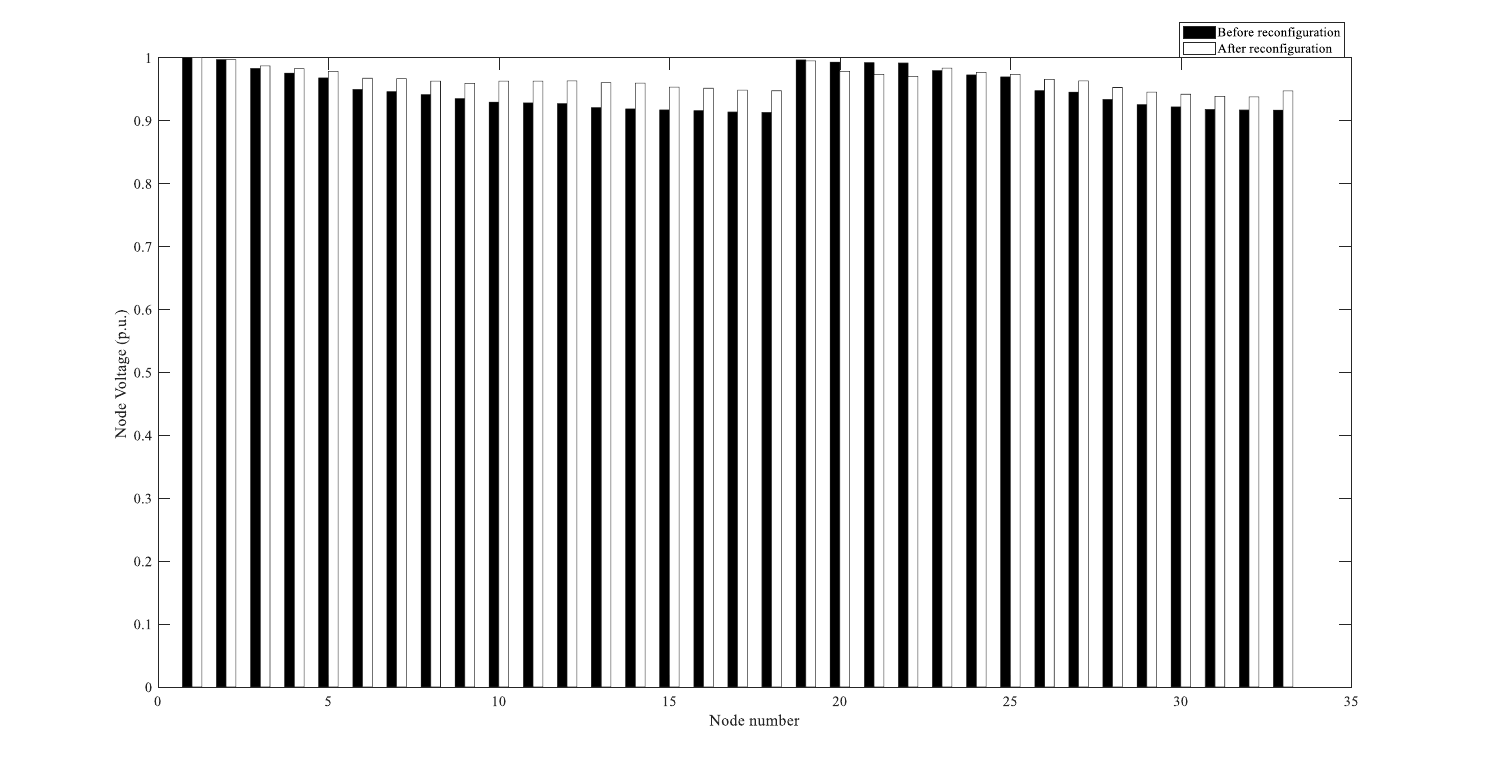}
    \caption{Voltage profile all nodes of 33-node system}
    \label{Voltage33}
\end{figure*}

For the determination of base case power loss, all the tie-lines are opened. This results in total active power loss of $202.3$ kW. The results obtained using the proposed method are compared with Harmony Search Algorithm (HSA) used by authors in \cite{6205640} and a heuristic method used by authors in \cite{7286576}. The comparison of results for $33$-node distribution test system is presented in Table \ref{tab:33node}. The voltage profile of $33$-node system before and after reconfiguration is shown in Fig. \ref{Voltage33}.
\begin{figure*}
    \hspace{-8ex}
    \includegraphics[scale=1.3]{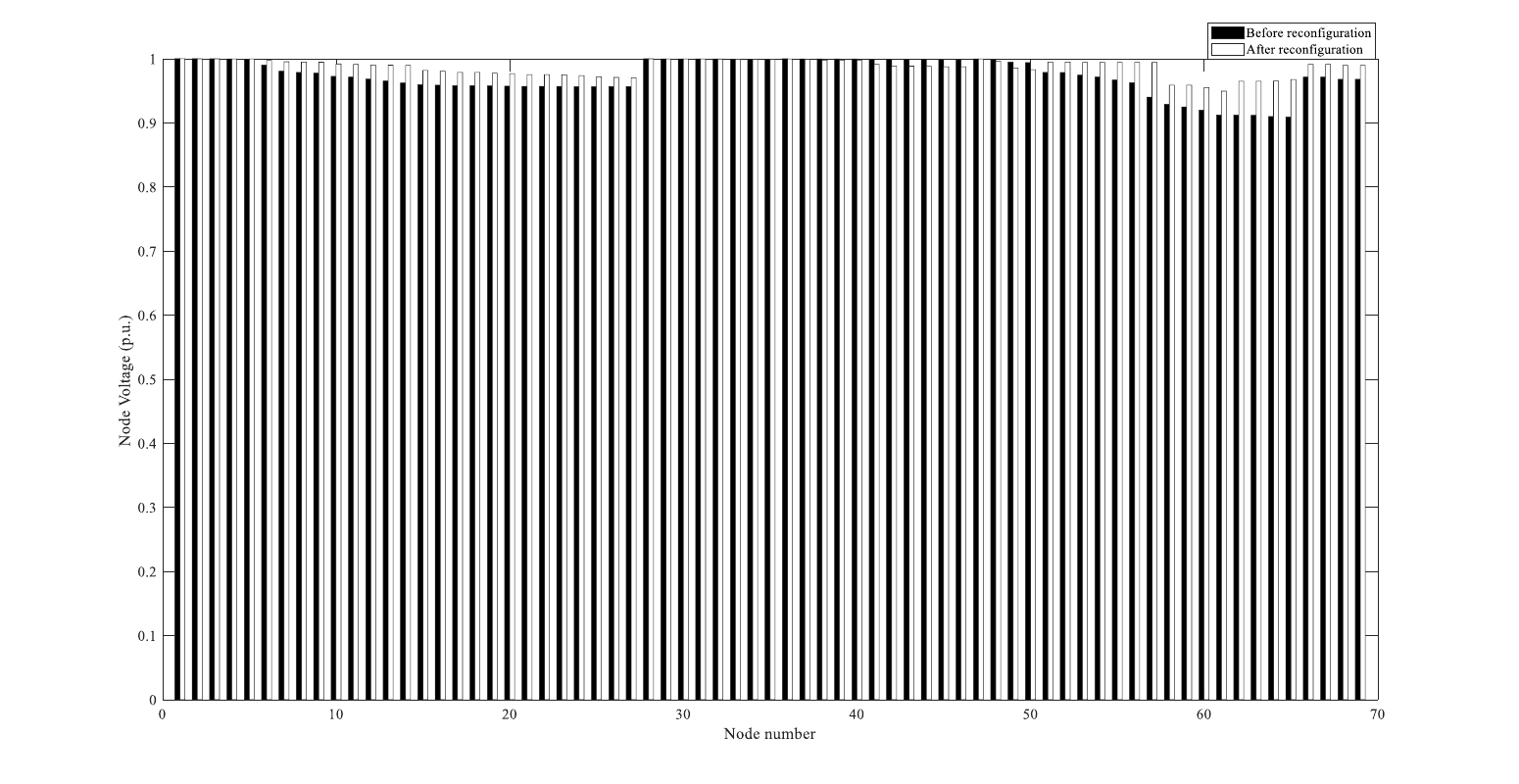}
    \caption{Voltage profile of all nodes of 69-node system}
    \label{Voltage69}
\end{figure*}
\begin{table}[h!]
\caption{Comparison of different methods for 33-node system\vspace{-1ex}}
    \centering
        \label{tab:33node}
    \begin{tabular}{c|c|c}
    \hline
        Methods  & Total active power & Opened branches \T \\
         & loss ($kW$) & after reconfiguration \T\\
        \hline
         Proposed & $139.1$  & $7,9,14,32,37$ \T\\
         HSA \cite{6205640} & $139.1$ & $7,9,14,32,37$ \T\\
         Heuristic Method \cite{7286576} & $140.5$ & $7,10,14,28,32$ \T\\
         \hline
    \end{tabular}
\end{table}

\begin{table}
\caption{Comparison of different methods for 69-node system\vspace{-1ex}}
    \centering
        \label{tab:69node}
    \begin{tabular}{c|c|c}
    \hline
        Methods  & Total active & Opened branches \T \\
         & power loss ($kW$) & after reconfiguration \T\\
        \hline
         Proposed & $96.20$  & $14,57,61,69,70$ \T\\
         HSA \cite{6205640} & $99.35$ & $13,18,56,61,69$ \T\\
         Heuristic Method \cite{7286576} & $99.59$ & $14,56,61,69,70$ \T\\
         Meta-heuristic \cite{SWARNKAR20111619} & $99.59$ & $14,56,61,69,70$ \T\\
         \hline
    \end{tabular}
\end{table}

The 69-node distribution test system is $12.66$ kV radial distribution system with $69$ nodes, $68$ branches and $5$ tie-lines (detail data is given in \cite{4302553}). Therefore, the total number of branches in this system is $73$. The $69$-node system is shown in Fig.\ref{69bus}, where all the branches (including tie-lines) are numbered from $1$ to $73$.

For the determination of base case power loss, all the tie-lines are opened. This results in total active power loss of $224.2$ kW. The results obtained using the proposed method are compared with a Harmony Search Algorithm (HSA) that has been used in \cite{6205640}, a heuristic method that has been used in \cite{7286576}, and a meta-heuristic method that has been used in \cite{SWARNKAR20111619}. The comparison of results for the $69$-node distribution test system is presented in Table \ref{tab:69node}. The voltage profile of the $69$-node system before and after the reconfiguration is shown in Fig. \ref{Voltage69}.

The comparison of results for different methods after the implementation of the proposed method on the $33$-node and $69$-node distribution test systems shows that the proposed method can determine the configuration that has lower active power loss compared to the other methods, which demonstrates its accuracy and efficiency.

\section{Conclusion}\label{conclusion}
This paper has proposed a spanning tree based genetic algorithm for optimal electric distribution system reconfiguration to minimize total active power losses. Spanning tree search algorithm was used to filter all invalid combinations of configurations that are generated during selection operation of genetic algorithm. The genetic algorithm was used to determine the optimal reconfiguration that reduces the total active power losses of distribution systems. Real (integer) numbers, instead of binary numbers, were used for the representation of individual chromosomes in the population as it reduces the dimensionality of GA search space. The proposed method was implemented on the $33$-node and $69$-node distribution test systems. The results obtained for both the test systems show that the method is accurate and efficient compared to existing methods.

\bibliographystyle{IEEEtran}
\bibliography{References.bib}

\begin{thebibliography}{10}
\providecommand{\url}[1]{#1}
\csname url@samestyle\endcsname
\providecommand{\newblock}{\relax}
\providecommand{\bibinfo}[2]{#2}
\providecommand{\BIBentrySTDinterwordspacing}{\spaceskip=0pt\relax}
\providecommand{\BIBentryALTinterwordstretchfactor}{4}
\providecommand{\BIBentryALTinterwordspacing}{\spaceskip=\fontdimen2\font plus
\BIBentryALTinterwordstretchfactor\fontdimen3\font minus
  \fontdimen4\font\relax}
\providecommand{\BIBforeignlanguage}[2]{{%
\expandafter\ifx\csname l@#1\endcsname\relax
\typeout{** WARNING: IEEEtran.bst: No hyphenation pattern has been}%
\typeout{** loaded for the language `#1'. Using the pattern for}%
\typeout{** the default language instead.}%
\else
\language=\csname l@#1\endcsname
\fi
#2}}
\providecommand{\BIBdecl}{\relax}
\BIBdecl

\bibitem{MerlinandBack}
{A. Merlin} and {H. Back}, ``Search for minimum-loss operating spanning tree
  configuration in an urban power distribution system,'' in \emph{Proc. 5th
  Power System Computation Conference}, 1975.

\bibitem{4113836}
C.~H. {Castro}, J.~B. {Bunch}, and T.~M. {Topka}, ``Generalized algorithms for
  distribution feeder deployment and sectionalizing,'' \emph{IEEE Transactions
  on Power Apparatus and Systems}, vol. PAS-99, no.~2, pp. 549--557, March
  1980.

\bibitem{CASTRO1985155}
C.~Castro and A.~FranÃ§a, ``Automatic power distribution reconfiguration
  algorithm including operating constraints,'' \emph{IFAC Proceedings Volumes},
  vol.~18, no.~7, pp. 155 -- 160, 1985.

\bibitem{6331584}
L.~S.~M. {Guedes}, A.~C. {Lisboa}, D.~A.~G. {Vieira}, and R.~R. {Saldanha}, ``A
  multiobjective heuristic for reconfiguration of the electrical radial
  network,'' \emph{IEEE Transactions on Power Delivery}, vol.~28, no.~1, pp.
  311--319, Jan 2013.

\bibitem{4302553}
J.~S. {Savier} and D.~{Das}, ``Impact of network reconfiguration on loss
  allocation of radial distribution systems,'' \emph{IEEE Transactions on Power
  Delivery}, vol.~22, no.~4, pp. 2473--2480, Oct 2007.

\bibitem{PES2020NB}
N.~Bhusal, M.~Gautam, and M.~Benidris, ``Sizing of movable energy resources for
  service restoration and reliability enhancement,'' in \emph{2020 IEEE Power
  Energy Society General Meeting (PESGM)}, Montreal, QC, Canada, Aug 2020, pp.
  1--5.

\bibitem{6205640}
R.~S. {Rao}, K.~{Ravindra}, K.~{Satish}, and S.~V.~L. {Narasimham}, ``Power
  loss minimization in distribution system using network reconfiguration in the
  presence of distributed generation,'' \emph{IEEE Transactions on Power
  Systems}, vol.~28, no.~1, pp. 317--325, Feb 2013.

\bibitem{SWARNKAR20111619}
A.~Swarnkar, N.~Gupta, and K.~Niazi, ``A novel codification for meta-heuristic
  techniques used in distribution network reconfiguration,'' \emph{Electric
  Power Systems Research}, vol.~81, no.~7, pp. 1619 -- 1626, 2011.

\bibitem{8316173}
I.~I. {Atteya}, H.~{Ashour}, N.~{Fahmi}, and D.~{Strickland}, ``Radial
  distribution network reconfiguration for power losses reduction using a
  modified particle swarm optimisation,'' \emph{CIRED - Open Access Proceedings
  Journal}, vol. 2017, no.~1, pp. 2505--2508, 2017.

\bibitem{7776824}
N.~C. {Koutsoukis}, D.~O. {Siagkas}, P.~S. {Georgilakis}, and N.~D.
  {Hatziargyriou}, ``Online reconfiguration of active distribution networks for
  maximum integration of distributed generation,'' \emph{IEEE Transactions on
  Automation Science and Engineering}, vol.~14, no.~2, pp. 437--448, April
  2017.

\bibitem{6344718}
D.~P. {Montoya} and J.~M. {Ramirez}, ``A minimal spanning tree algorithm for
  distribution networks configuration,'' in \emph{2012 IEEE Power and Energy
  Society General Meeting}, July 2012, pp. 1--7.

\bibitem{7867177}
{Lei Zhang}, {Kaoshe Zhang}, and {Gang Zhang}, ``Power distribution system
  reconfiguration based on genetic algorithm,'' in \emph{2016 IEEE Advanced
  Information Management, Communicates, Electronic and Automation Control
  Conference (IMCEC)}, Oct 2016, pp. 80--84.

\bibitem{holland1975}
J.~{Holland}, \emph{Adpatation in Natural and Artificial Systems}.\hskip 1em
  plus 0.5em minus 0.4em\relax University of Michigan Press, 1975.

\bibitem{Goldberg1989}
D.~{Goldberg}, \emph{Genetic algorithms in search, optimization and machine
  learning}.\hskip 1em plus 0.5em minus 0.4em\relax Addison-Wesley Publishing
  Company, Inc, 1989.

\bibitem{6826880}
S.~{Elsaiah}, M.~{Benidris}, and J.~{Mitra}, ``Analytical approach for
  placement and sizing of distributed generation on distribution systems,''
  \emph{IET Generation, Transmission Distribution}, vol.~8, no.~6, pp.
  1039--1049, June 2014.

\bibitem{7286576}
S.~{Elsaiah} and J.~{Mitra}, ``A method for minimum loss reconfiguration of
  radial distribution systems,'' in \emph{2015 IEEE Power Energy Society
  General Meeting}, July 2015, pp. 1--5.

\end{thebibliography}
\end{document}